\documentclass[twocolumn,showpacs,preprintnumbers,amsmath,amssymb]{revtex4}
\usepackage[dvipdfmx]{graphicx}
\usepackage{amsmath,amssymb}
\usepackage{bm}
\usepackage{color}

\newcommand{\Name}[1]{{\scshape #1},}
\newcommand{\Review}[1]{{\itshape #1,}}
\newcommand{\Vol}[1]{{\bfseries #1}}
\newcommand{\Year}[1]{(#1)}
\newcommand{\Page}[1]{\normalfont #1}
\newcommand{\Book}[1]{{\itshape #1}}
\newcommand{\REVIEW}[4]{\Review{#1} \Vol{#2} \Year{#3} \Page{#4}}
\newcommand{\etal}{\unskip\ \emph{et al.}}
\newcommand{\AND}{{\normalfont and} }

\newcommand{\nnmb}{\nonumber\\}

\newcommand{\braket}[3]{\left<{#1}\left|{#2}\right|{#3}\right>}

\begin{document}

\preprint{YITP-12-69}
\preprint{ICRR-Report 623-2012-12}

\author{Kazuyuki Sugimura$^1$}
\email{sugimura@yukawa.kyoto-u.ac.jp}
\author{Daisuke Yamauchi$^2$}
\email{yamauchi@icrr.u-tokyo.ac.jp}
\author{Misao Sasaki$^1$}
\email{misao@yukawa.kyoto-u.ac.jp}
\affiliation{%
$^1$Yukawa Institute for Theoretical Physics, Kyoto University, Kyoto, Japan\\
$^2$Institute for Cosmic Ray Research, University of Tokyo, Chiba, Japan
}%

\title{Non-Gaussian bubbles in the sky
  }

\date{\today}

\begin{abstract}
We point out a possible generation mechanism of non-Gaussian bubbles in the sky
due to bubble nucleation in the early universe.
We consider a curvaton scenario for inflation and assume
that the curvaton field $\phi$, whose energy density is subdominant during 
inflation but which is responsible for the curvature perturbation
of the universe, is coupled to another field $\sigma$ which undergoes false 
vacuum decay through quantum tunneling.
For this model, we compute the skewness of the curvaton 
fluctuations due to its interaction with $\sigma$ during tunneling,
that is, on the background of an instanton solution that describes false vacuum
decay. We find that the resulting skewness of the curvaton can become 
large in the spacetime region inside the bubble.
We then compute the corresponding skewness in the statistical
distribution of the cosmic microwave background (CMB)
temperature fluctuations. We find a non-vanishing skewness in a bubble-shaped
region in the sky. It can be large enough to be detected in the near future,
and if detected
it will bring us invaluable information about the physics in
the early universe.
\end{abstract}


\maketitle

\section{Introduction}
\label{sec:introduction}
Inflation, a stage of accelerated expansion in the very early universe, 
is now widely accepted as part of the standard evolutionary scenario 
of the universe.
On the other hand, many models of inflation have been proposed
but we are still far from being able to narrow down the possible models
sufficiently. Among those many models of inflation, much attention has 
been paid recently to the ones based on string 
theory~\cite{Kachru:2003aw},
which is considered to be a promising candidate for 
the ultimate unified theory.
In particular, it is of great interest if the string theory
landscape~\cite{Susskind:2003kw},
in which there are many local minima, or false vacua,
and the universe jumps from one minimum to another by quantum
tunneling, can be observationally tested~\cite{Yamauchi:2011qq,Yamamoto:1996qq}.
Quantum tunneling of a scalar field with gravity 
is usually treated with the Coleman-De Luccia (CDL) instanton
method~\cite{Coleman:1980aw}, where the evolution of a scalar field is 
described with an O(4)-symmetric instanton, which is a solution of
the Euclidean equations of motion.
Motivated by the string landscape,  inflation models 
with  multi-scalar fields and/or with tunneling are now keenly studied.
Extension of the CDL instanton method to a multi-scalar field system is already 
discussed in \cite{Sugimura:2011tk}.
In those studies, however, many inflation models have been proposed,
and now it is important to distinguish those models by observation.

Observations of the power spectrum of CMB temperature anisotropies 
have convinced us of the existence of an inflationary stage in the
very early universe.
On top of that, if a non-Gaussian feature such as skewness or bispectrum 
is detected in the CMB anisotropy,
it will have a strong impact on the physics of the early
universe~\cite{Komatsu:2010fb}.
In particular, since any single-field slow-roll inflation model
with canonical kinetic term predicts almost Gaussian 
fluctuations~\cite{Maldacena:2002vr},
any non-zero non-Gaussianity will exclude all these models.
Observations of non-Gaussianity use templates, 
such as local type~\cite{Komatsu:2001rj},\footnote{
In the view point of statistical distribution,
local type non-Gaussianity, which is local in the sense of 
generation mechanism, is homogeneous and isotropic.} 
equilateral type~\cite{Creminelli:2005hu},
and orthogonal type~\cite{Senatore:2009gt}, in order to increase 
the statistical significance.
However, since all of these templates assume statistical isotropy,
there may be anisotropic non-Gaussianities which 
may not have been detected by these templates.

In this letter, we study a multi-field model in which a nonlinear
interaction between two scalar fields, one of which being responsible
for the curvature perturbation of the universe (that is, for the
formation of the large scale structure) and the other
for quantum tunneling via a CDL instanton, 
induces an anisotropic non-Gaussianity.
To be specific, we introduce an inflaton field $\Phi$ that 
realizes slow-roll inflation,
a tunneling field $\sigma$ that governs the tunneling dynamics,
and a curvaton field $\phi$ that contributes to
the curvature perturbation of the 
universe~\cite{Lyth:2002my,Sasaki:2006kq}.

The inflaton dominates the energy density of the universe during inflation
but rapidly decays to radiation after the end of inflation,
On the other hand, 
the energy density of the curvaton field is negligible during inflation
but its decay is delayed after inflation so that it gradually begins to 
dominate the universe.

Approximating the universe during inflation by an exact de Sitter spacetime,
the inflaton behaves as a cosmic clock and determines
an appropriate time-slicing, namely, a spatially flat time-slicing of 
the de Sitter spacetime.
In this setup, assuming that the energy scale associated
with the tunneling field is much smaller than the energy scale
of inflation, the bubble nucleation can be well described by 
a single-field CDL instanton with no backreaction
to the geometry, that is, on the exact de Sitter
background~\cite{Coleman:1980aw,Yamamoto:1996qq}.

The curvaton $\phi$ is affected by the background bubble through 
a coupling with the tunneling field $\sigma$.
For simplicity and definiteness, we consider
a potential of the form, 
$V_\mathrm{int}(\sigma ,\phi )=\tilde\lambda (\sigma )\phi^3$\,.
Assuming that $V_\mathrm{int}(\sigma, \phi )$ is non-vanishing only 
at or inside the bubble wall,
we expect that $\phi$ may have a spatially localized,
bubble-shaped non-Gaussianity
due to the background bubble-shaped configuration of $\sigma$.
This leads to an anisotropic, bubble-shaped skewness
of the CMB temperature anisotropy.

This paper is organized as follows.
We first illustrate the background spacetime and the configuration
of the bubble.
Next, we briefly review a useful formalism for computing
the equal-time $N$-point functions, the tunneling in-in formalism,
and calculate the skewness of the curvaton fluctuations.
Then we demonstrate that a sky map of 
an anisotropic non-Gaussian parameter $f_{\rm NL}$
in our model.
Finally, we end with a few concluding remarks.

\section{Background Evolution}
\label{sec:bg_bubble}

\begin{figure}
\centering
  \includegraphics[width=6cm]{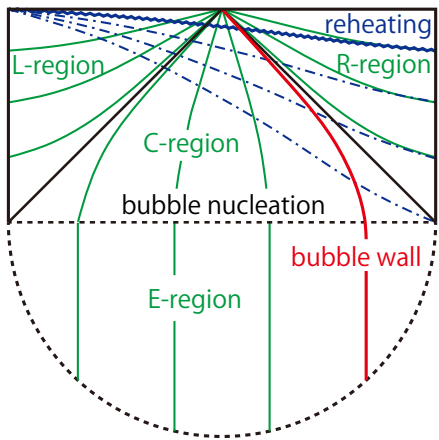}
\caption{Penrose-like diagram of bubble nucleating universe,
where a half of Euclidean region (bottom) and Lorentzian region after
 bubble nucleation (top)
are shown The blue dot-dashed lines are surfaces of constant cosmic time,
the green solid lines those of uniform tunneling field,
and the red solid line the location of the bubble wall.
}
\label{fig:penrose_ds}
\end{figure}

Let us start from a brief description of de Sitter 
spacetime~\cite{birrell-davies,Yamamoto:1996qq}.
A Lorentzian 4-dimensional de Sitter spacetime is O(4,1)-symmetric, 
which is obtained by analytical continuation
of an O(5)-symmetric 4-dimensional Euclidean sphere,
as illustrated in Fig.~\ref{fig:penrose_ds}.
Among various coordinatization of the spacetime,
the uniform tunneling field slicing is most appropriate to
see O(4)(O(3,1))-symmetry of the CDL instanton,
which corresponds to the time-slicing inside the bubble that
describes a spatially homogeneous and isotropic open universe.
For brevity, let us call it open slicing.
Open slicing in the C-region in Fig.~\ref{fig:penrose_ds}
is a time-like slicing,
an the metric in the C-region may be expressed as
\begin{align}
ds^2 &=\frac{1}{H^2}
 \Bigl(d\chi^2
 +\sin^2\chi\left( -d\tau^2
+\cosh^2\tau\,d\Omega_2^2\right)\Bigr)\,,
\label{openslicing}
\end{align}
where $-\infty<\tau<\infty$ and $0\leq\chi\leq\pi$, and
$d\Omega_2^2$ is the metric on the unit 2-sphere.
Theses coordinates can be extended to the other regions
of the spacetime 
by analytical continuation:
\begin{eqnarray}
&&\tau=r_\mathrm{R}+i\pi/2=r_\mathrm{L}-i\pi/2=ir_\mathrm{E}\,,
\cr
&&\chi=it_\mathrm{R}=-it_\mathrm{L}=t_\mathrm{E}\,,
\end{eqnarray}
where $(t_\mathrm{R},r_\mathrm{R})$, $(t_\mathrm{L},r_\mathrm{L})$ and
$(t_\mathrm{E},r_\mathrm{E})$ are the coordinates for
the R-, L-, and E-regions, respectively, in Fig.~\ref{fig:penrose_ds}.

The metric for the flat time-slicing is
\begin{eqnarray}
ds^2=-dt^2+a^2(t)\left(dr^2+r^2d\Omega_2^2\right)\,;
\ \,a(t)=H^{-1}e^{Ht}\,.
\label{flatslicing}
\end{eqnarray}
As mentioned in the introduction, the $t=const.$ slices are on which the
inflaton is uniform, and give the cosmic rest frame of our universe.

Let us briefly describe the bubble configuration of $\sigma$
under the thin-wall approximation.
The O(4)-symmetric CDL instanton for $\sigma$
can be described as a function of $\chi$,
which we denote by $\bar\sigma (\chi)$ hereafter.
We denote the wall radius by $R_{\rm W}$.
Hence
\begin{eqnarray}
\bar{\sigma}(\chi)=
\left\{
\begin{array}{ll}
\sigma_{\rm T}\quad\mbox{for}~&0\leq\chi<HR_{\rm W}\,,
\\
\sigma_{\rm F}\quad\mbox{for}~&HR_{\rm W}<\chi \leq \pi\,.
\end{array}
\right.
\end{eqnarray}
Though $\bar{\sigma}$ is homogeneous in the R- or L-regions,
the physical bubble radius $\ell_{\rm W}(t)$ on flat slices
increases as time goes on.
We take the origin $r=t=0$ of the flat slicing metric (\ref{flatslicing})
to be the center of the bubble at the time of nucleation.
Then the relations between the coordinates in the metrics~(\ref{openslicing})
 and (\ref{flatslicing}) are 
$\cos \chi = \cosh Ht - (1/2) e^{Ht}r^2$ and $\sin \chi\cosh\tau = e^{Ht} r$,
which gives $\ell_{\rm W}(t)=a(t)\sqrt{1+e^{-2Ht}-2e^{-Ht}\cos HR_{\rm W}}$.
As seen from this expression,
a bubble once nucleated with radius $\ell_{\rm W}(0)$ ($\approx R_{\rm W}$ 
for $HR_{\rm W}\ll 1$) expands to the Hubble horizon scale 
within one or two $e$-folds of time, $\Delta t\sim H^{-1}$,
and then expands comovingly as $\ell_{\rm W}(t)\sim a(t)$.

It should be noted that models with multiple nucleation is possible.
If those bubbles do not interact each other, we can take into
account the effect of all bubbles by summing up the effect of each bubble.
Hereafter, we consider a model with a single bubble for simplicity.

\section{Non-Gaussianity Generation}
\label{sec:ng-bubble-de}
\subsection{Effective Action on Instanton Background}
\label{sec:action}

Here we calculate the skewness in the curvaton fluctuations
on the single bubble background.
We consider the Lagrangian for the curvaton $\phi$ as
\begin{align}
\mathcal{L}_\phi
=-\sqrt{-g}
\left(\frac{1}{2}g^{\mu\nu}\partial_\mu\phi\partial_\nu\phi
+\frac{m^2}{2}\phi^2+V_\mathrm{int}(\sigma ,\phi )\right)\,,
\label{eq:5}
\end{align}
where $m^2\ll H^2$.
We expand the above by setting
$\phi=\phi_0+\delta\phi$,
where $\phi_0$ is a homogeneous classical part and $\delta\phi$ 
is a quantum fluctuation.
We assume that $\phi_0$ is approximately constant during inflation,
and we concentrate on the evolution of $\delta\phi$.
Further, for simplicity, we assume
$V_\mathrm{int}(\sigma,\phi)$
is non-vanishing only on the wall. Hence we approximate
it as $V_\mathrm{int}(\sigma,\phi)=\lambda\,\phi^3H\,\delta (\chi -HR_{\rm W})$
with $\lambda H=\int^{\pi}_0\mathrm{d}\chi\tilde{\lambda}(\bar{\sigma}(\chi))$\,.
Then the Lagrangian for $\delta\phi$
is given as $\mathcal{L}_{\delta\phi}=\mathcal{L}_0+\mathcal{L}_I$,
where $\mathcal{L}_0$ is free-part and $\mathcal{L}_I$ is interaction-part,
\begin{align}
\mathcal{L}_0&=-\sqrt{-g}\left(\frac{1}{2}
g^{\mu\nu}\partial_\mu\delta\phi\,\partial_\nu\delta\phi\,
+\frac{1}{2}\left( m^2+\delta m^2\right)\delta\phi^2
\right),\nnmb
\mathcal{L}_I
&=-\sqrt{-g}\,\lambda H\delta\left(\chi-HR_{\rm W}\right)\delta\phi^3\,,
\label{eq:1}
\end{align}
where $\delta m^2\equiv 6\lambda H\phi_0\delta (\chi-HR_{\rm W})$.

\subsection{Quantum Field Theory on the Instanton Background}
\label{sec:qft-inst-backgr}

To calculate the quantum fluctuations on the instanton background,
we extend the in-in formalism to Euclidean spacetime, which
we call the tunneling in-in formalism.
This formalism is based on the WKB analysis of a tunneling wave function 
for free theory~\cite{Tanaka:1993ez,Yamamoto:1993mp}
to the case with nonlinear interactions.
A detailed derivation will be given
elsewhere~\cite{preparation}.\footnote{A similar 
formalism is used in \cite{Park:2011ty}.
However, the motivation of~\cite{Park:2011ty} was 
to show a theoretical relation called the FLRW-CFT correspondence
and hence it is quite different from ours.}

\begin{figure}
\centering
  \includegraphics[width=5cm]{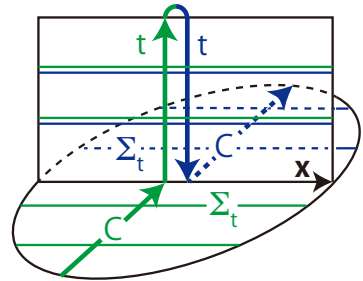}
\caption{Same figure as Fig.~\ref{fig:penrose_ds}, but 
with all domains of integration given in eq.~\eqref{eq:11}.
}
\label{fig:path_pen}
\end{figure}

The tunneling in-in formalism tells us that 
the $N$-point function of $\delta\phi$ is given by 
\begin{align}
&\Big<\delta\phi(x_1)\delta\phi(x_2)\cdots\delta\phi(x_N)\Big>\nnmb
=& \frac{\braket{0}
{P\,\delta\phi(x_1)\delta\phi(x_2)\cdots\delta\phi(x_N)
e^{i\int_{C\times \Sigma_{t}}dtd^3x\,\mathcal{L}_I}}{0}}
{\braket{0}
{P
e^{i\int_{C\times \Sigma_{t}}dtd^3x\,\mathcal{L}_I}
}{0}},
\label{eq:11}
\end{align}
where the tunneling in-in path $C$ and $t=const.$ surfaces 
$\Sigma_t$ are as shown in Fig.~\ref{fig:path_pen}.
The first half of $C$ (the arrowed green line in Fig.~\ref{fig:path_pen})
goes from one end of the Euclidean region
to future infinity in the Lorentzian region
through the bubble nucleation surface.
The second half of $C$ (the arrowed blue line in Fig.~\ref{fig:path_pen})
goes back from future infinity through the nucleation surface
to the other end of the Euclidean region.
The slicing $C\times\Sigma_{t}$ covers the whole Euclidean region
and the future half of the Lorentzian region twice.
In the Lorentzian region any $\Sigma_{t}$ is a Cauchy surface.
The operator $P$ in eq.~\eqref{eq:11} is the path-ordering
operator.
In the Lorentzian region, $P$ reduces to the
time-ordering and anti-time-ordering operators $T$ and $\bar{T}$
on the first and second halves of $C$, respectively. 
It should be noted that eq.~\eqref{eq:11} is independent of 
the choice of coordinates.

Eq.~\eqref{eq:11} is evaluated in the same way as in
the usual perturbation theory.
After expanding the interaction part perturbatively,
operators are transformed to products of the
free correlation function 
$G(x,x')=\braket{0}{\delta\phi(x)\delta\phi(x')}{0}$ using Wick's theorem. 
We note that $G(x,x')$ is not the correlation function for 
the Bunch-Davies vacuum due to the non-trivial nature of $\mathcal{L}_0$.
It may be obtained by studying the ``evolution'' of the mode 
functions in the Euclidean space~\cite{Yamamoto:1996qq}.

While $G(x,x')$ is a single-valued function when $x$ and $x'$ are
in space-like separation,
when they are in time-like separation,
$T$ or $\bar{T}$ in the expression singles out
the Feynman or anti-Feynman propagator, respectively.
A branch of $G(x,x')$, when $x'$ is in the Euclidean region,
is determined by analyticity on $x'$ along $C$.
It may be noted that by this way of choosing branches the expression in
eq.~\eqref{eq:11} is equivalent to that
obtained by analytical continuation of Euclidean quantum
 field theory~\cite{Park:2011ty},
and O(4)(O(3,1)) symmetry of the result is guaranteed.

\subsection{Skewness from Bubble Nucleation}
\label{sec:evaluation}

Now we evaluate the skewness $\langle\delta\phi^3(x)\rangle$ 
by substituting $\mathcal{L}_{\delta\phi}$ in eq.~\eqref{eq:1}
to the tunneling in-in formula in eq.~\eqref{eq:11}.
In the following calculations, we approximate $G(x,x')$ 
by that of the Bunch-Davies vacuum, by neglecting corrections 
due to the non-trivial part of $\mathcal{L}_0$ in eq.~\eqref{eq:1}.
This effect may affect the value of the skewness,
but since it only induces a statistically homogeneous non-Gaussianity,
we simply ignore it here.

The free correlation function $G(x,x')$ of the Bunch-Davies vacuum is 
given by~\cite{birrell-davies}
\begin{align}
\!\!\!\!\!\! G(x,x')
\!&=\!\frac{H^2(\frac{1}{4}-v^2)}{16\pi\cos\pi v}\,
_2F_1\!\left[
\begin{matrix}
\frac{3}{2}+v,\frac{3}{2}-v\\
2
\end{matrix}
;\frac{1\!+\!Z(x,x')}{2}\right],
\label{eq:31}
\end{align}
where $v= \sqrt{9/4-m^2/H^2}$
and $Z(x,x')=\cos (Hd(x,x'))$, where $d(x,x')$ is the geodesic distance 
between $x$ and $x'$. We have $Z>1$, $Z=1$ and $Z<1$, 
for a timelike (imaginary $d$), null ($d=0$), and spacelike (real $d$)
separation, respectively.
As mentioned before, a branch of $G(x,x')=G(Z(x,x'))$ should be specified
when $Z>1$ and it can be done by adding small imaginary part $\pm i\epsilon$
to $Z$ as $G(Z(x,x')\pm i\epsilon)$\,.

\begin{figure}
\centering
  \includegraphics[width=7cm]{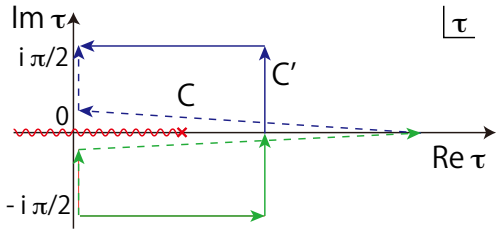}
\caption{The structure of integrand and integration path in eq.~\eqref{eq:42}.
The wavy line is the branch cut,
the dotted line is the original path $C$, while the
solid line is a detoured path $C'$ used for numerical evaluation.
}
\label{fig:des_inin}
\end{figure}

We assume $\lambda \ll 1$ to avoid possible strong coupling problems.
To leading order in $\lambda$, we obtain
\begin{align}
 \Big<\delta\phi^3 (x)\Big>
&= -i \lambda H^{-3}
 \int_{C}\mathrm{d}\tau'
\int_0^\pi\mathrm{d}\chi'
 \int\mathrm{d}\Omega'\nnmb
 \times&\sin^3\chi'\sinh^2\tau'\,
 \delta\left(\chi' -HR_{\rm W}\right)\left( G(x,x')\right)^3\,.
 \label{eq:42}
\end{align}
where 
\begin{eqnarray}
C:\ -i\,\frac{\pi}{2}\to 0-i\,\epsilon \to \infty
\to 0+i\,\epsilon \to i\,\frac{\pi}{2}\,,
\end{eqnarray}
as shown in Fig.~\ref{fig:des_inin}.
Among the domains of integral, $\tau'\in(-i\,\pi/2,+i\,\pi/2)$ 
and $\tau'\in(0\pm i\,\epsilon, \infty )$
correspond to the E- and C-regions, respectively.
Integration in the R- or L-regions is unnecessary
because $\mathcal{L}_I$ vanishes there.
A small imaginary part in $\tau'\in(0\pm i\epsilon, \infty )$ 
gives a small imaginary part to $Z(x,x')$ to select the Feynman or anti-Feynman
propagator.

Evaluation of eq.~\eqref{eq:42} is straightforward in principle,
but it is significantly simplified if we use the O(4)-symmetry
of the $N$-point functions in the tunneling in-in formalism.
Just for illustration, let us consider the case when
$x$ is in the R-region. In this case, computing 
$\langle\delta\phi^3 (x)\rangle$ at $r_\mathrm{R}=0$, where
integration over $\chi$ and $\Omega$ is trivial,
is enough since the value at any other point in the R-region can 
be known from the O(4)-symmetry.

After integrating over $\chi$ and $\Omega$, the integration over
 $\tau$ along $C$ remains.
Since the original path $C$ passes near the singularity, 
it makes the integral difficult to evaluate numerically.
To avoid it, we deform the path $C$ to another path $C'$
without crossing the branch cut or the poles of the integrand
as in Fig.~\ref{fig:des_inin}.
The resulting $\langle\delta\phi^3 (x)\rangle$ is 
obtained in terms of the open-slice coordinates,
but the expression in terms of the flat-slice coordinates
can be easily found by the coordinate transformation.

\begin{figure}
 \centering
\includegraphics[width=7cm]{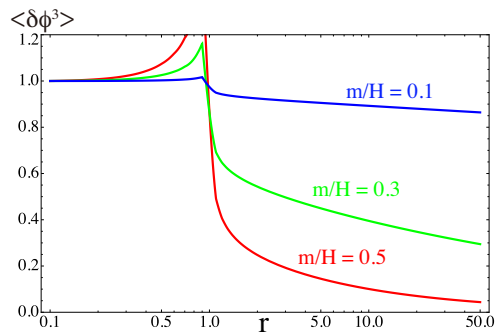}
\caption{$\langle\delta\phi^3(r,t)\rangle$ normalized by the value at
the center as a function of $r$ at $Ht=50$.}
\label{fig:phi3_tf0}
\end{figure}

\begin{figure}
 \centering
\includegraphics[width=7cm]{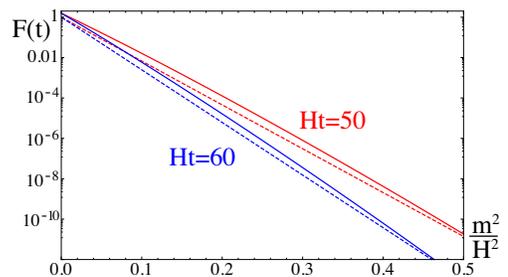}
\caption{A normalized skewness as a function of time, $F(t)$.
The solid lines are the result of numerical evaluation and
the dashed lines are analytical approximation, $e^{-\frac{m^2}{H^2}Ht}$.
}
\label{fig:f_mh}
\end{figure}

Fig.~\ref{fig:phi3_tf0} shows the skewness on flat slices at time $t$
as a function of the radial coordinate $r$, 
$\langle\delta\phi^3 (r,t)\rangle /\langle\delta\phi^3 (0,t)\rangle$,
for $Ht=50$ and $HR_{\rm W}=0.2\pi$ for several values of $m/H$.
To see the dependence of the skewness on the model parameters, 
we introduce a normalized skewness $F(t)$ 
at the center of the bubble by
\begin{eqnarray}
\langle\delta\phi^3 (0,t)\rangle
=4\pi\lambda\,\sin^3(HR_{\rm W})\left(\frac{3H^2}{8\pi^2m^2}\right)^3H^3F(t)\,.
\end{eqnarray}
The result is shown in Fig.~\ref{fig:f_mh}.
From this, we see that $F\approx e^{-\frac{m^2}{H^2}Ht}$.

We see that the skewness is large inside the bubble 
and it decreases as one moves away from the bubble,
and its radial dependence is stronger for larger $m/H$.
These features can be understood from the fact that
$G(x,x')$ is roughly given by $(3H^4/8\pi^2m^2)|1-Z|^{-m^2/(3H^2)}$ 
for $m/H\ll1$ and $Z$ is not very close to 1. 
The apparent divergent behavior at the bubble wall comes from
the UV divergence in $G(x,x')$.
But, this divergence disappears as soon as the finite resolution of observations
or renormalization of the theory is taken into account.


\section{Non-Gaussian Bubbles in the Sky}
\label{sec:ng-bubble-slow}

Now we study signatures of our model in the
CMB anisotropy. In our model, the evolution of the universe after the 
inflaton and the tunneling field decay,
is exactly the same as the ordinary curvaton 
scenario~\cite{Lyth:2002my}.
Here, we estimate the curvature perturbation after the curvaton decay
by using the sudden-decay approximation~\cite{Sasaki:2006kq}.

After inflation the curvaton starts to roll down the potential and undergoes
damped oscillations when the Hubble parameter becomes smaller than $m$.
The curvaton energy density behaves like a pressureless matter during
this stage of damped oscillations.
At the time $t=t_\mathrm{curv}$ when the curvatons decay,
the energy density of the universe consists of
that of radiation $\rho_r$ generated from the decay of the inflatons
and that of the curvaton $\rho_\phi$.
The contribution of each component to the total curvature perturbation
of the universe may be conveniently expressed in terms of
$\zeta_r$ and $\zeta_\phi$, which are defined as the curvature perturbations 
on the uniform energy density slices of the radiation and the curvaton, respectively.

It is known that each of them is separately conserved on superhorizon scales 
at $t<t_\mathrm{curv}$. As for $\zeta_\phi$, it may be evaluated
by the energy density fluctuations on the uniform inflaton field slice
at the end of inflation, $\delta\rho_\phi/\rho_\phi$, as
\begin{eqnarray}
\zeta_\phi=\frac{1}{3}\left.\frac{\delta\rho_\phi}{\rho_\phi}\right|_{t_{\rm e}}
=\frac{1}{3}\left.\left(2\,\frac{\delta\phi}{\phi_0}
 +\frac{\delta\phi^2}{\phi_0^2}\right)\right|_{t_{\rm e}},
\end{eqnarray}
where $t_{\rm e}$ is the time at the end of inflation.

After the curvaton decay, the total curvature perturbation
on the slices of uniform total energy density, $\zeta$,
is conserved on superhorizon scales. It is given by
\begin{eqnarray}
\zeta =(1-r_\phi)\zeta_r+r_\phi\zeta_\phi\,,
\end{eqnarray}
where $r_\phi=3\rho_\phi/(4\rho_r+3\rho_\phi)|_{t_{\rm curv}}$.

For the moment, we ignore the contribution from $\delta\phi^2/\phi_0^2$
and concentrate on the skewness generated by the nonlinear 
interaction with the tunneling field.
Then we have
\begin{align}
\langle\zeta^3(r)\rangle\approx
 r_\phi^3\langle\zeta_\phi^3(r,t_\mathrm{e})\rangle
\approx\frac{8}{27}r_\phi^3
\frac{\langle\delta\phi^3(r,t_\mathrm{e})\rangle}{\phi_0^3}
\,,
\label{eq:7}
\end{align}
where $\langle\delta\phi^3\rangle$ is given by Eq.~\eqref{eq:42}.
An important feature is that the skewness of the curvature perturbation
depends on the position of the bubble due to the spatial dependence
of $\langle\delta\phi^3\rangle$.

To proceed, we focus only on the large angular scale CMB anisotropy for 
which the Sachs-Wolfe effect dominates. In this regime, 
the CMB temperature anisotropy can be written in terms of
the curvature perturbation as 
$(\delta T/T)(\hat{\bm n})=(1/5)\,\zeta ({\bm x}_0+r_*\hat{\bm n},t_*)$\,,
where ${\bm x}_0$ is the comoving position of the observer 
measured from the bubble center,
$\hat{\bm n}=(\sin\theta\cos\varphi ,\sin\theta\sin\varphi ,\cos\theta )$
is the directional cosine of the sky seen by the observer,
$r_*$ is the comoving distance from the observer to the last scattering surface
and $t_*$ is the time at the last scattering surface.
Then, defining the non-Gaussianity parameter $f_{\rm NL}$ as
$(3/5)f_{\rm NL}(\hat{\bm n})
\equiv\langle\zeta^3 ({\bm x}_O+r_*\hat{\bm n},t_*)\rangle
/\langle\zeta^2\rangle^2$,
we have
\begin{align}
f_{\rm NL}(\hat{\bm n})
\approx\frac{40r_\phi^3}{81\langle\zeta^2\rangle^2}
\frac{\langle\delta\phi^3 (|{\bm x}_O+r_*\hat{\bm n}|, t_*)\rangle}{\phi_0^3}
\,.
\label{eq:2}
\end{align}
One explicitly sees that the dependence on the position of the bubble center
breaks the statistical isotropy.

For completeness, let us discuss the additional contributions to $f_{\rm NL}$ 
which we have ignored. 
The contribution of $\zeta_r$ to non-Gaussianity is known to be very small
as long as the vacuum is in the Bunch-Davies vacuum~\cite{Maldacena:2002vr}.
A deviation from the Bunch-Davies vacuum may give rise to a non-zero
non-Gaussianity. It is expected to be statistically homogeneous and isotropic
but scale-dependent (in the Fourier space). It may be detected by 
the templates of the equilateral
or orthogonal types. 
The contribution from the part of $\zeta_\phi$ quadratic in $\delta\phi/\phi_0$
gives rise to a local type non-Gaussianity which is again statistically
homogeneous and isotropic, and which may be detected by the
squeezed type templates.

This contribution from the nonlinearity of $\delta\phi/\phi_0$
in $\zeta_\phi$ may be estimated as follows.
Roughly speaking,
$\langle\delta\phi^4\rangle\approx\langle\delta\phi^2\rangle^2\approx (H/2\pi )^4$
for $m/H\ll 1$,
and the contribution to $\langle\zeta_\phi^3\rangle$ is about 
$(4/9) (H/2\pi \phi_0)^4$.
Thus, the condition that $f_{\rm NL}$ is dominated by the nonlinear interaction
with the bubble is given as
\begin{eqnarray*}
\frac{4}{9}\left(\frac{H}{2\pi\phi_0}\right)^4
\ll
\frac{32\pi}{27}\lambda\, \sin^3(HR_{\rm W})
\left(\frac{3H^2H}{8\pi^2m^2\phi_0}\right)^3
F(t_\mathrm{e})\,.
\end{eqnarray*}
This is satisfied in an example we compute below.

\begin{figure}
\centering
\includegraphics[width=7cm]{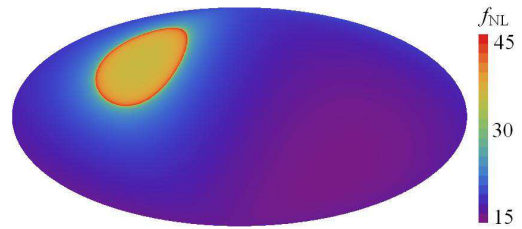}
\caption{Non-Gaussianity map in the CMB sky,
for the model parameters given in the text.
Colors correspond to the values of $f_{\rm NL}$.
}
\label{fig:skymap}
\end{figure}

For illustration, we plot $f_{\rm NL}(\hat{\bm n})$ in the CMB sky 
in Fig.~\ref{fig:skymap}.
The parameters are
$m/H=0.3$, $HR_{\rm W}=0.2\pi$ $H/\phi_0=0.001$ $\lambda=0.005$,
$r_\phi=0.1$, $|{\bm x}_0|=r_*=2$, $Ht_\mathrm{e}=50$.
For the variance $\langle\zeta^2\rangle$ we simply impose 
the observational result, 
$\langle\zeta^2\rangle =A_\zeta^2\equiv 6.25\times 10^{-10}$~\cite{Komatsu:2010fb},
assuming that it is dominated by $\zeta_r$.\footnote{
In this case, it may not be appropriate to call $\phi$ a curvaton,
since it never dominates the curvature perturbations. Nevertheless, 
we call it a curvaton for notational convenience.}
Since $\langle\delta\phi^2\rangle\approx (H/2\pi )^2$ when $m/H\ll 1$,
this assumption is justified if $(2r_\phi /3)^2(H/2\pi\phi_0)^2 \ll A_\zeta^2$,
which is marginally satisfied in the above example.

The typical value of $f_{\rm NL}$ at the center of the bubble
when the parameters satisfy the above conditions is estimated as
\begin{align}
f_{\rm NL}^{\mathrm{(cen)}}\!
\approx \frac{3\!\times\! 10^{-4}\,\lambda\, r_\phi^3\,\sin^3(\!HR_{\rm W}\!)}
{A_\zeta^4\,\exp\left({(\frac{m}{H})^2Ht_\mathrm{e}}\right)}
\left(\!\frac{H}{m}\!\right)^{\!6}\!\left(\!\frac{H}{\phi_0}\!\right)^{\!3}\!\!
\,.
\end{align}
This gives $f_{\rm NL}^{\mathrm{(cen)}}\approx 15$
for the parameters given above and
agrees with the result in Fig.~\ref{fig:skymap} within 
the errors shown in Fig.~\ref{fig:f_mh}. From this estimation, 
we find that the resultant $f_{\rm NL}$ 
is rather sensitive to the values of $m/H$ and $H/\phi_0$.
It is enhanced for smaller $m/H$ and larger $H/\phi_0$.

\section{Conclusion}
\label{sec:conclusion}

In this paper, we calculated the skewness in 
the CMB temperature anisotropy
in a model with bubble nucleation during inflation,
motivated by the string theory landscape.
We considered bubble nucleation in the curvaton scenario of inflation
in which the curvaton vacuum fluctuations are 
affected by the bubble nucleation through interaction with 
the tunneling field. The calculation was
done by extending the in-in formalism to the instanton 
background~\cite{preparation}.
We found that there can be spatially localized, bubble-shaped skewness
which is large inside the bubble.

As far as we know, bubble-shaped non-Gaussianities have
not been studied carefully yet in observation.
So it seems interesting to look for such a non-Gaussianity
already in the current observational data.
As pointed out by Komatsu and Spergel~\cite{Komatsu:2001rj},
the bispectrum, corresponding to the 3-point function,
contains much more information than the single value of the
skewness. Thus, analysis beyond skewness may improve the observability 
of bubble-shaped non-Gaussianities.
We hope to come back to this issue in a future publication.

Since the string theory landscape gives a strong motivation for
inflation models with bubble nucleation,
studies of such inflation models may be regarded as testing
string theory using the universe as a laboratory.
In any case, if any signature of bubble nucleation during inflation is found
in observation, it gives a huge impact on the physics
of the early universe, including string theory.

Finally, we note that in the evaluation of the skewness
we neglected the effect of deviations from the Bunch-Davies vacuum.
This may affect details of our result, though generic features are 
expected to remain the same. This effect can be evaluated by 
studying the evolution of mode function on the
instanton background~\cite{Yamamoto:1996qq}. We plan to
come back to this issue in the near future.

\begin{acknowledgments}
KS thanks T.~Tanaka, Y.~Korai and K.~Iwaki for useful discussions and 
valuable comments.
This work was supported in part by Monbukagaku-sho 
Grant-in-Aid for the Global COE programs, 
``The Next Generation of Physics, Spun from Universality 
and Emergence'' at Kyoto University.
This work was also supported in part by JSPS Grant-in-Aid for
 Scientific Research (A) No.~21244033,
and by Grant-in-Aid for Creative Scientific Research No.~19GS0219.
KS was supported by Grant-in-Aid for JSPS Fellows
No.~23-3437. 
\end{acknowledgments}


\end{document}